\documentclass[prd,twocolumn,showpacs,preprintnumbers,superscriptaddress,floatfix,nofootinbib]{revtex4-1}
\usepackage{subfigure}
\usepackage{amsfonts}
\usepackage{graphicx,,booktabs}
\usepackage{amssymb}
\usepackage{amsmath,txfonts,ulem}
\usepackage{graphicx}
\usepackage{dcolumn}
\usepackage{multirow}
\usepackage{color}
\usepackage{bm}
\usepackage{ulem}
\usepackage[colorlinks,
            citecolor=blue,
            anchorcolor=green,
            menucolor=orange,
            linkcolor=red,
            filecolor=red,
            runcolor=pink,
            urlcolor=blue,
            frenchlinks=red]{hyperref}

\allowdisplaybreaks[4]

\begin{document}

\title{Exploration of trace anomaly contribution to proton mass based on light vector meson photoproduction}
    \author{Chen Dong}
	\email{dongphysics@yeah.net}
	\affiliation{Department of physics, Lanzhou University of Technology, Lanzhou 730050, China}
	
	\author{Jiyuan Zhang}
	\email{jyuanzhang@yeah.net}
	\affiliation{Department of physics, Lanzhou University of Technology, Lanzhou 730050, China}
	
	\author{Jingxuan Bu}
	\affiliation{Department of physics, Lanzhou University of Technology,
		Lanzhou 730050, China}
		
	\author{Huifang Zhou}
	\affiliation{Department of physics, Lanzhou University of Technology,	
	Lanzhou 730050, China}

	\author{Xiao-Yun Wang}
	\email{Corresponding author:xywang@lut.edu.cn}
	\affiliation{Department of physics, Lanzhou University of Technology,
		Lanzhou 730050, China}
	\affiliation{Lanzhou Center for Theoretical Physics, Key Laboratory of Theoretical Physics of Gansu Province, Lanzhou University, Lanzhou, Gansu 730000, China}

\begin{abstract}
In the quantum chromodynamics, the mass source of the proton is decomposed into four parts by the energy momentum tensor : quark energy term, gluon energy term, quark mass term and trace anomaly term. And the trace anomaly term is the most crucial contribution for studying the internal structure of the proton. In this work, under the definition of the vector meson dominant model, the trace anomaly contribution of the proton is extracted from the experimental datas of light vector mesons $\rho$, $\omega$ and $\phi$ photoproduction at the near-threshold, which are ($1.15\pm0.08$)$\%$, ($2.70\pm0.04$)$\%$ and ($5.58\pm0.25$)$\%$, respectively. Eventually, the average trace anomaly contribution of the proton is ($4.36\pm0.40$) $\%$,  which only account for a small fraction of the total proton mass.
The result of this work will provide theoretical information support for further study of proton internal structure.

\end{abstract}


\maketitle

\section{Introduction}
The emergence of the quantum chromodynamics Energy Momentum Tensor (EMT) provides a direct computational standard for understanding various properties of hadrons \cite{Polyakov:2018zvc}, such as mass, spin \cite{Ji:1996ek}, angular momentum \cite{Lorce:2017wkb,Granados:2019zjw,Lorce:2019zyw}, mechanical properties\cite{Polyakov:2018zvc,Wang:2022uch,Kharzeev:2021qkd,Kou:2021qdc,Polyakov:2002wz,Lorce:2018egm,Wang:2022vhr,Lorce:2019zyw}, etc. All these attributes are the essential properties of hadrons, which are valuable to further exploring the internal structure of hadrons. As the most basic and stable hadron, the study of the internal structure of the proton is one of the most meaningful topics to be discussed \cite{Kharzeev:2021qkd,Metz:2020vxd}.

The proton is composed of quarks and gluons, with a total mass of $M_{N}=0.938$ GeV. However, the distribution of quarks and gluons inside the proton and their respective percentages of the total mass has been an ambiguous question. The first mass decomposition of the proton originated in 1995, Ji \cite{Ji:1994av,Ji:1995sv} decomposed the mass source of the proton into four terms based on the EMT $T^{\mu\nu}$ of hadronic states : the quark energy term $M_{q}$, the gluon energy term $M_{g}$, the quark mass term $M_{m}$ and the trace anomaly term $M_{a}$. Also related to EMT focusing on the decomposition of $T^{00}$ comes from the work of Lorcé \cite{Lorce:2017xzd} and Hatta \cite{Hatta:2018sqd}. Lorcé's \cite{Lorce:2017xzd} decomposition consists of two schemes: a two-term decomposition and a four-term decomposition.  Hatta \cite{Hatta:2018sqd} primarily concentrates on the regularization scheme to prove that the trace anomaly solely comes from the gluon momentum energy tensor, while the quark mass term just comes from the quark energy tensor. Beyond that, a revisiting of the mass decomposition is developed by A. Metz et al. in 2020 \cite{Metz:2020vxd}, who reviewed the previous similarities and differences in the origin of the proton mass and concluded that any decomposition at most exists two independent terms. This is because the existence of the forward limit leads to the EMT matrix element having two independent form factors \cite{Lorce:2017xzd}. The most striking point is that using independent operators \cite{DUNE:2021tad} in their four-term decomposition computations yields the trace anomaly term $M_{a} =0$, which is different from the results of other mass decompositions. The above research investigates the source of protons in terms of momentum energy tensor and whether other ways exist to enable us to comprehend the decomposition of the proton.

A novel approach in Ref. \cite{Wang:2019mza,Kou:2021bez} is proposed to relate the differential cross sections of vector meson photoproduction to the trace anomaly parameter $b$ in $M_{a}$ with the vector meson dominate (VMD) model. The trace anomaly contribution of the proton is studied employing $\phi$ and $J/\psi$ photoproduction data from the LEPS Collaboration \cite{LEPS:2005hax},  SAPHIR \cite{Barth:2003bq} and GlueX experiment \cite{GlueX:2019mkq} at JLab, respectively.
A new report \cite{Duran:2022xag} recently measured a new set of the differential data of $J/\psi$ photoproduction at $E\in[9.10,10.60]$ GeV and extracted the trace anomaly contribution of the proton with VMD. 
One finds that the trace anomaly contribution extracted from the heavy meson photoproduction data varies rapidly with energy, indicating that the trace anomaly is very sensitive to the energy and has a strong dependence.
This causes considerable confusion in the conclusive determination of the trace anomaly contribution of the proton. 
Therefore, it is necessary to study the trace anomaly contribution in terms of the cross section of light vector mesons photoproduction at the near-threshold.

As the most familiar light vector mesons, $\rho$, $\omega$ and $\phi$ have accumulated considerable experimental data \cite{Dey:2014tfa,LEPS:2005hax,Strakovsky:2020uqs,rho-data,Barth:2003kv,Wu:2005wf} during years of development. In this work, based on the photoproduction experimental data of the three mesons, the trace anomaly contribution of the proton is extracted by VMD model and analyzed as a function of $R$ at the near-threshold range. Here, $R$ is proportional to the center of mass energy $W$, representing the ratio of the final momentum $|{\bf p}_3|$ to the initial momentum $|{\bf p}_1|$. In addition, since the two gluon exchange model is reliable in predicting the mass radius of the proton \cite{Wang:2022uch}, the trace anomaly contribution of the proton is expected based on the two gluon exchange model. Combined with the result extracted from the photoproduction data of the three vector mesons at the near-threshold, the trace anomaly contribution of the proton is systematically discussed.

The outline of the paper is organized as follows. The expression of the mass decomposition of the proton, the vector meson dominate and the two gluon exchange models are described in Sec. \ref{sec2},  along with a detailed discussion of trace anomaly. The results of trace anomaly coherence are presented in Sec. \ref{sec3}. A simple summary and discussion is presented  in Sec. \ref{sec4}.

\section{Formalism} \label{sec2}
\subsection{Mass decomposition of the proton}
Starting with the EMT, $T^{\mu \nu}$ is decomposed into the traceless term $\Bar{T}_{\mu \nu}$ and trace term $\hat{T}^{\mu \nu}$ \cite{Ji:1995sv,Ji:1994av},
\begin{equation}
    T^{\mu \nu}=\Bar{T}^{\mu \nu}+\hat{T}^{\mu \nu}\label{eq:1.1}
\end{equation}
where the traceless $\Bar{T}_{\mu \nu}$ contains contributions from gluon and quark, and the trace $\hat{T}^{\mu \nu}$ includes contributions from quark mass and trace anomaly. A more detailed decomposition is,
\begin{equation}
\begin{aligned}
&\Bar{T}_{\mu \nu}=\Bar{T}^{\mu \nu}_{q}+\Bar{T}^{\mu \nu}_{g}, \\
&\hat{T}^{\mu \nu}=\hat{T}^{\mu \nu}_{m}+\hat{T}^{\mu \nu}_{a}.
\end{aligned}\label{eq:1.2}
\end{equation}
Then, the EMT is expressed as a sum of four terms,
\begin{equation}
    T^{\mu \nu}=\Bar{T}^{\mu \nu}_{q}+\Bar{T}^{\mu \nu}_{g}+\hat{T}^{\mu \nu}_{m}+\hat{T}^{\mu \nu}_{a}.
    \label{eq:1.3}
\end{equation}

Introduce the matrix element of the energy momentum tensor in the proton state $|P\rangle$, then the left side of Eq. (\ref{eq:1.2}) is normalized as
\begin{equation}
\begin{aligned}
&\left\langle P\left|\bar{T}^{\mu \nu}\right| P\right\rangle=\left(P^{\mu} P^{\nu}-\frac{1}{4} g^{\mu \nu} M^{2}_{N}\right) / M_{N}, \\
&\left\langle P\left|\hat{T}^{\mu \nu}\right| P\right\rangle=\frac{1}{4} g^{\mu \nu} M_{N} .
\end{aligned} 
\label{eq:1.4}
\end{equation}
where $g^{\mu \nu}$ is metric tensor of space. The normalization requires Lorentzian invariant while renormalization scales independently. In the rest frame, the total energy of the mass of the whole EMT can be defined by the Hamiltonian form, which is 
\begin{equation}
\begin{aligned}
&\left\langle\bar{T}^{00}\right\rangle \mid _{P=0}=\frac{3}{4} M_{N}, \\
&\left\langle\hat{T}^{00}\right\rangle \mid _{P=0}=\frac{1}{4} M_{N}.
\end{aligned} 
\label{eq:1.5}
\end{equation}
That is, the trace part is $25\%$ of the proton total mass $M_{N}$, which means $\left\langle P\left|\hat{T}^{00}_{m}\right| P\right\rangle+\left\langle P\left|\hat{T}^{00}_{a}\right| P\right\rangle=M_{N}/4$. Under the Hamiltonian operator, Eq. (\ref{eq:1.3}) can be reformulated as \cite{Lorce:2017xzd},
\begin{equation}
    \mathcal{H}=\mathcal{H}_{q}+\mathcal{H}_{g}+\mathcal{H}_{m}+\mathcal{H}_{a}.
    \label{eq:1.6}
\end{equation}
Hence, in the rest frame, the mass decomposition of the proton is derived from Eq. \ref{eq:1.6},
\begin{equation}
    M_{N}=M_{q}+M_{g}+M_{m}+M_{a}.
    \label{eq:1.7}
\end{equation}
The four mass components with Hamiltonian form are parameterized as,
\begin{equation}
	\begin{aligned}
		M_{q} &=\frac{3}{4}\left(a-\frac{b}{1+\gamma_{m}}\right) M_{N} \\
		M_{g} &=\frac{3}{4}(1-a) M_{N} \\
		M_{m} &=\frac{4+\gamma_{m}}{4\left(1+\gamma_{m}\right)} b M_{N} \\
		M_{a} &=\frac{1}{4}\left(1-b\right) M_{N}.
	\end{aligned}\label{eq:1.8}
\end{equation}
where $b$ is trace anomaly parameter, $a$ is the momentum fraction of the quarks and $\gamma_{m}$ is the anomalous dimension of the quark mass \cite{Hatta:2018sqd,Buras:1979yt,Baikov:2014qja}. In the chiral limit and low energy scattering, $b=0$, the proton's mass is entirely due to the contribution of gluon quantum fluctuations. In this case, $M_{m}=0$ and $M_{a}=M_{N}/4$.

\subsection{Vector meson dominate}

In Eq. (\ref{eq:1.8}), the fourth term is needed to compute the trace anomaly contribution $M_{a}/M_{N}$. This item contains a trace anomaly parameter $b$, which is the key to determining the contribution of trace anomaly. Therefore, the vector mesons dominate (VMD) model is developed to calculate parameter $b$ and $M_{a}/M_{N}$.

The differential cross section at $t=t_{min}$ for $\gamma N$$ \rightarrow$$V N$ process can be expressed with the VMD model,
\begin{equation}
	\begin{aligned}
		\left.\frac{d \sigma_{\gamma N \rightarrow V N}}{d t}\right|_{t=t_{min}}&=\left.\frac{3 \Gamma_{V \rightarrow e^{+} e^{-}}}{\alpha_{em} m_{V}}\left(\frac{{\bf p}_3}{{\bf p}_1}\right)^{2} \frac{d \sigma_{V N \rightarrow V N}}{d t}\right|_{t=t_{min}}\\
		&=\left.\frac{3 \Gamma_{V \rightarrow e^{+} e^{-}}}{\alpha_{em} m_{V}} R^{2} \frac{d \sigma_{V p \rightarrow V p}}{d t}\right|_{t=t_{min}},
	\end{aligned} 
	\label{eq:4}
\end{equation}
with
\begin{equation}
	{\bf p}_1=\frac{W^{2}-m_{N}^{2}}{2 W}, \label{eq:5}
\end{equation}
\begin{equation}
	{\bf p}_3=\sqrt{\frac{\left[W^{2}-\left(m_{V}+m_{N}\right)^{2}\right]\left[W^{2}-\left(m_{V}-m_{N}\right)^{2}\right]}{4 W^{2}}}, \label{eq:6}
\end{equation}
where $\alpha_{em}=1/137$ is the fine coupling constant, $\Gamma_{V \rightarrow e^{+} e^{-}}$ is the lepton decay width, $M_{N}$ is the mass of the proton, $m_{V}$ is the mass of the light meson ($\rho$, $\omega$ and $\phi$), and $R=|{\bf p}_3|/|{\bf p}_1|$ represents the ratio of the final momentum to the initial momentum. The $\left.\frac{d \sigma_{V N \rightarrow V N}}{d t}\right|_{t=t_{min}}$ is the elastic scattering in Eq. (\ref{eq:4}),
\begin{equation}
	\left.\frac{d \sigma_{V N \rightarrow V N}}{d t}\right|_{t=t_{min}}=\frac{1}{64 \pi} \frac{1}{m_{V}^{2}\left(\lambda^{2}-m_{N}^{2}\right)}\left|F_{V N}\right|^{2},  \label{eq:7}
\end{equation}
where $\lambda=(W^{2}-m^{2}_{V}-m^{2}_{N})/(2m_{V})$ is the nucleon energy \cite{Kharzeev:1998bz}. In the low energy region, the elastic scattering amplitude $F_{V N}$ of the particle is denoted as \cite{Kharzeev:1995ij},
\begin{equation}
	F_{V N}=r_{0}^{3} d_{2} \frac{2 \pi^{2}}{27} 2 M_{N}^{2}\left(1-b\right),  \label{eq:8}
\end{equation} 
with the Bohr radius of meson and Wilson coefficient \cite{Kou:2021bez,Kharzeev:1995ij,Kharzeev:1996tw},
\begin{equation}
r_{0}=\left(\frac{4}{3 \alpha_{s}}\right) \frac{1}{m_{q}},
\end{equation}
\begin{equation}
d_{n}^{(1 S)}=\left(\frac{32}{N_{c}}\right)^{2} \sqrt{\pi} \frac{\Gamma\left(n+\frac{5}{2}\right)}{\Gamma(n+5)},
\end{equation}
where $m_{q}$ is the constituent quark, $\alpha_{s}$ is the strong coupling constant and $b$ is trace anomaly parameter. Stay away from the chiral limit, the contribution of QCD trace anomaly to proton mass can be represented by the factor $1-b$. By the fourth term of Eq. (\ref{eq:1.8}), Eq. (\ref{eq:8}) becomes
\begin{equation}
	F_{V N}=r_{0}^{3} d_{2} \frac{2 \pi^{2}}{27} 2 M_{N}^{2} 4 \frac{M_{a}}{M_{N}}. \label{eq:8-1}
\end{equation} 

The total cross section can be obtained by integrating the differential cross section Eq. (\ref{eq:4}) from $t_{min}$ to $t_{max}$. When the centre of mass energy $W$ approaches the threshold, $t_{min}$$\rightarrow$$t_{max}$ , the total corss section has the integral \cite{Pentchev:2020kao},
\begin{equation}
    \sigma_{\gamma N \rightarrow V N}(W)=4 |{\bf p}_1| |{\bf p}_3|\frac{d \sigma_{\gamma N \rightarrow V N}}{d t} \label{eq:14}
\end{equation}
where $\Delta=|t_{max}-t_{min}|=4 |{\bf p}_1| |{\bf p}_3|$ \cite{Pentchev:2020kao}.
From the above statement, we can calculate the trace anomaly contribution $M_{a}/M_{N}$ from the total and differential cross sections of vector mesons.
\subsection{Two gluon exchange model}
\begin{figure}[htbp]
	\begin{center}
		\includegraphics[scale=0.36]{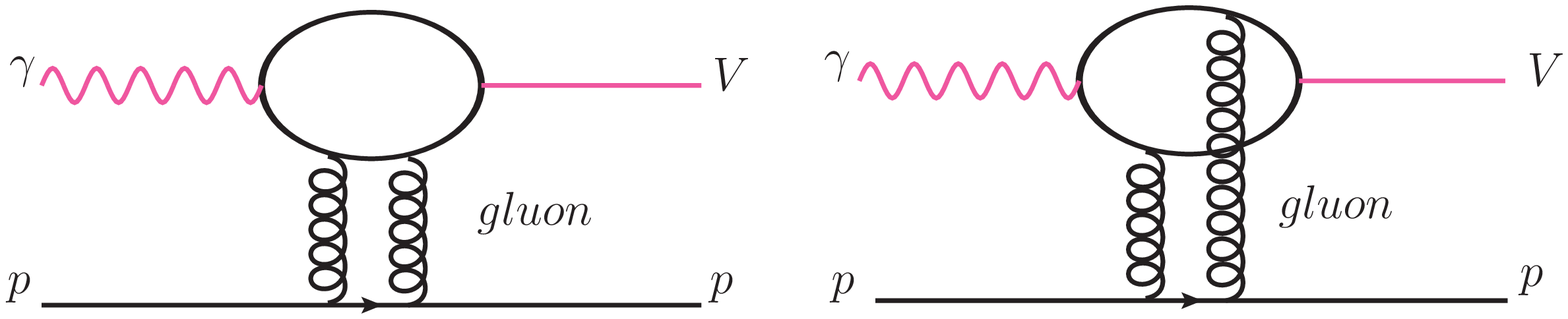}
		\caption{The Feynman diagrams of the two gluon exchange model for vector mesons photoproduction.}  \label{fig:two-gluon}
	\end{center}
\end{figure}

Studying the photoproduction reaction of vector mesons can not only provide insight into the behaviour of the photoproduction cross section of vector mesons but also explore the related properties of the proton, such as the source of mass, mass radius and mechanical properties of the proton \cite{Polyakov:2018zvc,Wang:2022uch,Kharzeev:2021qkd,Kou:2021qdc,Polyakov:2002wz,Lorce:2018egm,Wang:2022vhr}.
The process can be described as splitting photons into $q\bar{q}$ which exchanges the two gluons to scatter the initial proton.  Eventually, the $q\bar{q}$ pair hadrons form the final vector meson, as shown in Fig. \ref{fig:two-gluon}. This is the two gluon exchange model based on perturbation QCD, which can be directly related to the gluon distribution function $xg(x,m_{V}^{2})$. Due to the existing partial distribution functions such as CTEQ6M \cite{Goloskokov:2007zb}, GRV98 \cite{Gluck:1998xa}, NNPDF \cite{Ball:2011uy}
, CJ15 \cite{Owens:2012bv,Accardi:2016qay}, and IMParton16 \cite{Wang:2016sfq}. It isn't easy to interpret the relevant photoproduction experimental data of vector mesons at the near-threshold. Therefore, the parameterized gluon distribution function $xg(x,m_{V}^{2})$$=$$A_{0}x^{A_{1}}(1-x)^{A_{2}}$ \cite{Zeng:2020coc,Pumplin:2002vw} is introduced in this work. Here, $A_{0}$, $A_{1}$ and $A_{2}$ are free parameters.
Therefore, combined with $xg(x,m_{V}^{2})$, the differential cross section of the two gluon exchange model is \cite{Sibirtsev:2004ca}
\begin{equation}
\frac{d \sigma}{d t}=\frac{\pi^{3} \Gamma_{e^{+} e^{-}} \alpha_{s}}{6 \alpha m_{q}^{5}}\left[x g\left(x, m_{V}^{2}\right)\right]^{2} \exp \left(b_{0} t\right),
\label{eq:3.1}
\end{equation}
where $b_{0}$ is the slope which can be determined by experimental data. The total cross section of vector meson photoproduction is expressed by integrating Eq. (\ref{eq:3.1}) from $t_{min}$ to $t_{max}$, 
\begin{equation}
    \sigma(W)=\int_{t_{min}}^{t_{max}}\frac{d \sigma}{d t}d t  \label{eq:3.2}
\end{equation}
Here, $t_{min}$ ($t_{max}$) is expressed as \cite{Zeng:2020coc,peskin}
\begin{equation}
    t_{min}(t_{max})=(\frac{m_{1}^2-m_{3}^2-m_{2}^2+m_{4}^2}{2W})^{2}-(p_{1cm}\mp p_{3cm})^{2} \label{eq:3.3}
\end{equation}
where $p_{icm}$$=$$\sqrt{E_{icm}^{2}-m_{i}^{2}}$ ($i=1,3$), $E_{1cm}$$=$$(W^{2}+m_{1}^{2}-m_{2}^{2})/(2W)$ and $E_{3cm}$$=$$(W^{2}+m_{3}^{2}-m_{4}^{2})/(2W)$.
\section{RESULTS} \label{sec3}
In our previous work \cite{Wang:2022uch,Wang:2022tuc}, the mass radius of the proton and the scattering length of $\phi$-$N$ were predicted with the two gluon exchange model, which is in good agreement with those extracted directly from the experimental data. This phenomenon indicates that the two gluon exchange model is reliable for studying the physical properties of the proton. Therefore, in this work, we concentrate on the cross section of $\phi$ photoproduction at the near-threshold predicted by the two gluon exchange model to extract the trace anomaly contribution $M_{a}/M_{N}$ of the proton. The relevant parameters for $\gamma p$$\rightarrow$$\phi p$ based on the two gluon exchange model are list in Tab. \ref{tab:table0}. In addition, $\rho$ and $\omega$ are not composed of pure $s\bar{s}$, but a mixture of $ud$ quarks. So the trace anomaly contributions are extracted directly from the experiment data of $\rho$ and $\omega$ meson \cite{rho-data,Barth:2003kv,Wu:2005wf}. Some parameters related to $\rho$, $\omega$ and $\phi$ are listed in Tab. \ref{tab:table1}.

\begin{table} \small
	\caption{\label{tab:table0} 
	The slope $b_{0}$ and parameters $A_{0}, A_{1}, A_{2}$ for the two gluon exchange model.} 
	\begin{ruledtabular}
		\begin{tabular}{cccc}
			$A_{0}$&$A_{1}$&$A_{2}$&$b_{0}$ (GeV$^{-2}$) \\
			\hline
			$0.36 \pm 0.04$&$-0.055 \pm 0.003$ & $0.12 \pm 0.03$ &$3.60\pm0.04$\\
		\end{tabular}
	\end{ruledtabular}
\end{table}

\begin{table}\small
\caption{\label{tab:table1}The lepton decay width $\Gamma_{e^{+}e^{-}}$, the constituent quark mass $m_{q}$, meson mass $m_{V}$ and the strong coupling parameter $\alpha_{s}$ for $\rho$, $\omega$ and $\phi$.}
	\begin{ruledtabular}
		\begin{tabular}{ccccc}
			Meson  &$\Gamma_{e^{+}e^{-}}$ (keV)&$m_{q}$ (GeV) &$m_{V}$ (GeV)&$\alpha_{s}$ \\
			\hline
			$\rho$&$7.04$ \cite{Mathieu:2018xyc}&$0.330$ \cite{Zhao:1998rt}&$0.770$&$0.439$ \cite{Ganbold:2010bu} \\
			\hline
			 $\omega$&$0.60$ \cite{Mathieu:2018xyc}&$0.330$ \cite{Zhao:1998rt}&$0.782$&$0.460$ \cite{Ganbold:2010bu}\\
			 \hline
			$\phi$&$1.27$ \cite{Strakovsky:2020uqs}&$0.486$ \cite{Kou:2021bez}&$1.019$&$0.701$ \cite{Kou:2021bez}\\
			
		\end{tabular}
	\end{ruledtabular}
\end{table}

\begin{figure}[htbp]
	\begin{center}
		\includegraphics[scale=0.43]{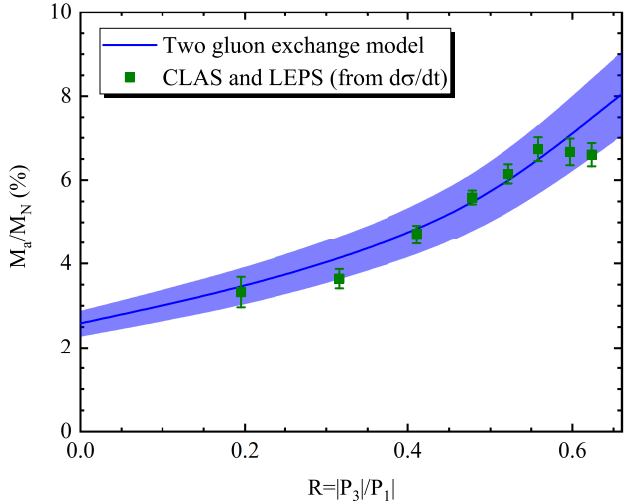}
		\caption{ The function of the trace anomaly contribution with $R$ based on the differential cross section of $\phi$ predicted by the two gluon exchange model (the blue line). The olive-green squares are extracted from the CLAS \cite{Dey:2014tfa} and LEPS \cite{LEPS:2005hax}. Compared with the result obtained by heavy vector meson photoproduction, the extraction of $M_{a}/M_{N}$ from the $\phi$ photoproduction is relatively flat with the change of energy.}  \label{fig:phiweifen}
	\end{center}
\end{figure}

Tab. \ref{tab:table2} lists the $M_{a}/M_{N}$ extracted directly from the experiment data of $\phi$ photoproduction from the CLAS \cite{Dey:2014tfa} and LEPS Collaboration \cite{LEPS:2005hax}, which shows an overall upward trend with increasing the centre of mass energy $W$. The average trace anomaly contribution $\sqrt{\left\langle (M_{a}/M_{N})^{2}\right\rangle}$ is calculated as ($5.58\pm0.25$)$\%$. At $W=1.98$ GeV,  $M_{a}/M_{N}=(5.28\pm2.50$)$\%$ in Ref.  \cite{Kou:2021bez} extracted from the differential cross section of $\phi$ photoproduction \cite{LEPS:2005hax} is significantly larger. This is because $t=t_{min}$ is adopted in this paper, while $t=0$ is selected  in Ref. \cite{Kou:2021bez}. $M_{a}/M_{N}$ as a function of $R$  based on the two gluon exchange model (the blue line) is shown in Fig. \ref{fig:phiweifen} which is consistent with the result extracted from the CLAS \cite{Dey:2014tfa} and LEPS data \cite{LEPS:2005hax}, and the $\sqrt{\left\langle M_{a}/M_{N})^{2}\right\rangle}=(6.03\pm0.76)$$\%$. Here, the interval of $R$ is $[0,0.66]$.

\begin{table}\small
\caption{\label{tab:table2} The trace anomaly contribution $M_{a}/M_{N}$ from the CLAS \cite{Dey:2014tfa} and LEPS datas \cite{LEPS:2005hax} of meson $\phi$.The average is ($5.58\pm0.25$)$\%$.}
	\begin{ruledtabular}
		\begin{tabular}{cccc}
			$W$ (GeV)  &$1.98$ &$2.02$&$2.07$ \\
			\hline
			$M_{a}/M_{N}$ ($\%$) &$3.32 \pm 0.36$&$3.64 \pm 0.23$&$4.70 \pm 0.20$\\
			\hline
			\hline
			 $W$ (GeV)&$2.12$&$2.16$&$2.20$\\
			 \hline
			$M_{a}/M_{N}$ ($\%$) &$5.58 \pm 0.17$&$6.14 \pm 0.23$&$6.74 \pm 0.29$\\
			\hline
			\hline
			$W$ (GeV)&$2.25$&$2.29$&-\\
			\hline
			$M_{a}/M_{N}$ ($\%$) &$6.67\pm0.32$&$6.60\pm0.28$&-
		\end{tabular}
	\end{ruledtabular}
\end{table}

\begin{figure}[htbp]
	\begin{center}
		\includegraphics[scale=0.43]{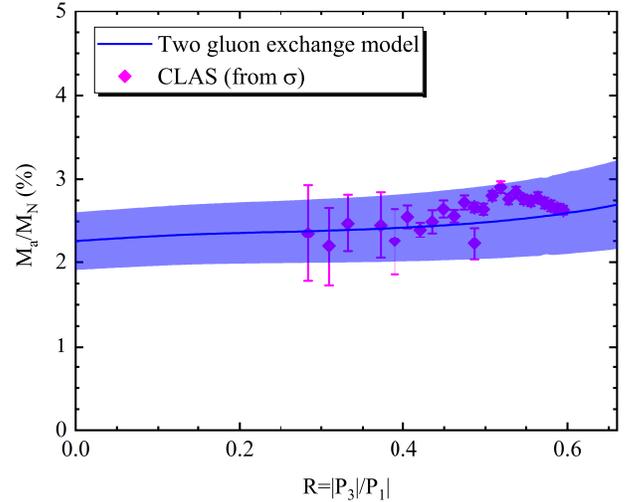}
		\caption{ The trend of the trace anomaly contribution with $R$ based on total cross section of $\phi$ predicted by the two gluon exchange model (the blue line).  The magenta diamond is extracted from the CLAS data \cite{Strakovsky:2020uqs}.}  \label{fig:phitot}
	\end{center}
\end{figure}

Unlike the trace anomaly contribution obtained from the differential cross section, $M_{a}/M_{N}$ extracted from the total cross section of $\phi$ photoproduction grows relatively gently with $R$, as shown in Fig. \ref{fig:phitot}.
The $\sqrt{\left\langle (M_{a}/M_{N})^{2}\right\rangle}=(2.51\pm0.45$)$\%$ is from the two gluon exchange model and $\sqrt{\left\langle (M_{a}/M_{N})^{2}\right\rangle}=(2.60\pm0.15$)$\%$ is extracted from the CLAS data \cite{Strakovsky:2020uqs}. 
Relating Fig. \ref{fig:phiweifen} and \ref{fig:phitot}, at the point closest to the threshold, the $M_{a}/M_{N}$ extracted from the total and differential cross sections of $\phi$ photoproduction predicted by the two gluon exchange model are adjacent, which takes into account the error bars. Moreover, the average value calculated from the total cross section at $R\in[0,0.66]$ is almost identical to the $M_{a}/M_{N}$ calculated from differential cross section at the threshold $R=0$ (the $M_{a}/M_{N}$ is ($2.57\pm0.30$)$\%$). This indicates that the trace anomaly extracted from the total cross section is the minimum of the trace anomaly extracted from the differential cross section.

\begin{table}\small
\caption{\label{tab:rho} The trace anomaly contribution from the differential experimental data of $\rho$ \cite{rho-data,Wu:2005wf}, where $W=1.79$, $1.82$, $1.87$, $1.94$ GeV are from Ref. \cite{rho-data} and $W=1.92$ GeV is from Ref. \cite{Wu:2005wf}.The average is $(1.15\pm0.08)$$\%$.}
	\begin{ruledtabular}
		\begin{tabular}{cccc}
			$W$ (GeV)  &$1.79$ &$1.82$&$1.87$ \\
			\hline
			$M_{a}/M_{N}$ ($\%$) &$0.81 \pm 0.08$&$1.10 \pm 0.06$&$1.14 \pm 0.07$\\
			\hline
			\hline
			 $W$ (GeV)&$1.94$&$1.92$&-\\
			 \hline
			$M_{a}/M_{N}$ ($\%$) &$1.24 \pm 0.08$&$1.38 \pm 0.10$&-\\
			
		\end{tabular}
	\end{ruledtabular}
\end{table}

\begin{table}\small
\caption{\label{tab:omega} The trace anomaly contribution from the differential experimental data of $\omega$ \cite{Barth:2003kv}. The average is $(2.70\pm0.04)$$\%$.}
\centering
\begin{ruledtabular}
\centering
\begin{tabular}{cccc}
W (GeV) & $M_{a}/M_{N}$ ($\%$) & W (GeV) & $M_{a}/M_{N}$ ($\%$) \\
$1.73$  & $1.30\pm0.03$ & $1.98$  & $2.60\pm0.04$ \\
$1.74$  & $1.66\pm0.03$ & $2.01$  & $2.79\pm0.04$ \\
$1.75$  & $1.85\pm0.03$ & $2.04$  & $2.83\pm0.03$ \\
$1.76$  & $1.87\pm0.03$ & $2.09$  & $3.08\pm0.04$ \\
$1.78$  & $1.94\pm0.02$ & $2.13$  & $3.42\pm0.04$ \\
$1.81$  & $2.04\pm0.02$ & $2.17$  & $3.32\pm0.04$ \\
$1.83$  & $1.99\pm0.02$ & $2.22$  & $2.32\pm0.05$ \\
$1.86$  & $2.08\pm0.03$ & $2.26$  & $3.40\pm0.05$ \\
$1.88$  & $2.15\pm0.03$ & $2.30$  & $3.68\pm0.06$ \\
$1.91$  & $2.21\pm0.03$ & $2.34$  & $3.47\pm0.06$ \\
$1.93$  & $2.34\pm0.03$ & $2.38$  & $4.02\pm0.07$ \\
$1.96$  & $2.52\pm0.04$ &    -     &    -          
\end{tabular}

\end{ruledtabular}
\end{table}

\begin{figure}[htbp]
	\begin{center}
		\includegraphics[scale=0.43]{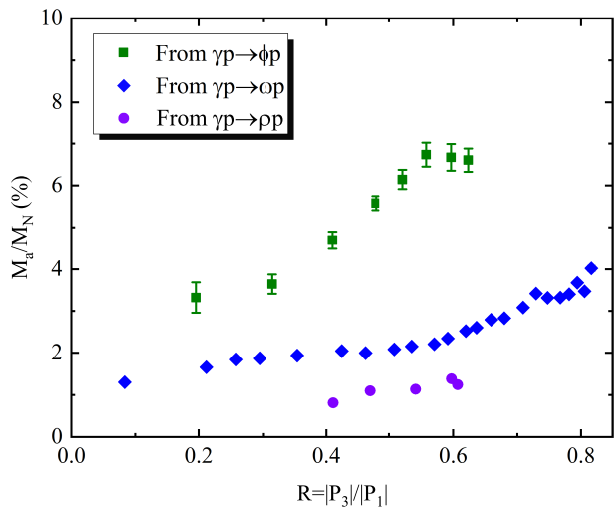}
		\caption{The trace anomaly contributions $M_{a}/M_{N}$ as a function of $R$ extracted from the differential experimental data of three light vector mesons photoproduction \cite{Dey:2014tfa,LEPS:2005hax,rho-data,Barth:2003kv,Wu:2005wf}. The olive green square is the result from $\gamma p$$\rightarrow$$\phi p$ from the CLAS \cite{Dey:2014tfa} and LEPS datas \cite{LEPS:2005hax}. The blue diamond is extracted from the data of $\omega$ derived by SAPHIR Collaboration\cite{Barth:2003kv}. 
		The purpl circle represents the $M_{a}/M_{N}$ extracted from the $\rho$ experimental data developed by SAPHIR Collaboration \cite{rho-data,Wu:2005wf}.
		}  \label{fig:three ma}
	\end{center}
\end{figure}

\begin{table}[]
\begin{ruledtabular}
\caption{\label{tab:compare} The $M_{a}/M_{N}$ extracted from the differential cross section of $\phi$, $\omega$ and $\rho$. The root mean square is ($4.36\pm0.40$)$\%$.}
\begin{tabular}{c|cc}
 \multirow{2}{*}{ Meson } & \multicolumn{2}{c}{$M_{a} / M_{N}(\%)$} \\
\cline { 2 - 3 } & Extraction from experimental & Two gluon exchange model \\
\hline$\phi$ & $5.58 \pm 0.25$ & $6.03 \pm 0.76$ \\
\hline$\omega$ & $2.70 \pm 0.04$ & $-$ \\
\hline$\rho$ & $1.15 \pm 0.10$ & $-$ \\
\end{tabular}
\end{ruledtabular}
\end{table}

A recent literature \cite{Duran:2022xag} has measured a new set of the differential cross sections of $J/\psi$ photoproduction and extracted the trace anomaly contribution. The results indicate that the $M_{a}/M_{N}$ increases with the energy $W$,  which is consistent with the results obtained by the differential cross section of $\phi$ photoproduction. This phenomenon shows that the trace anomaly is very sensitive and dependent on energy. However, it does not suggest that the consequence extracted from the total cross section of $\phi$ is incorrect. It's just because the trace anomaly contribution from the differential cross section is more reasonable to demonstrate the growing relationship.

$\rho$ and $\omega$ are two special vector mesons, and the $\gamma p$$\rightarrow$$\rho/\omega p$ process cannot be wholly established with the two gluon exchange model.
Based on the differential experimental cross section data of $\rho$ \cite{rho-data,Wu:2005wf} at $W\in[1.79,1.94]$ and $\omega$ \cite{Barth:2003kv} at $W\in[1.73,2.38]$ measured in the SAPHIR Collaboration \cite{rho-data,Barth:2003kv,Wu:2005wf}, the trace anomaly contributions extracted directly with the VMD model are listed in Tab. \ref{tab:rho} and \ref{tab:omega}. Near the threshold, $M_{a}/M_{N}$ obtained from $\rho$ and $\omega$ photoproduction data is tiny, with average values of ($1.16\pm0.07$)$\%$ and ($2.70\pm0.04$)$\%$, respectively. In order to apparently observe the trace anomaly relationship extracted from the $\rho$, $\omega$ and $\phi$ photoproduction cross section, $W$ is transformed into the corresponding $R$ here, which is shown in Fig. \ref{fig:three ma}.
With $R$ increases, the $M_{a}/M_{N}$ extracted from different vector mesons has an upward trend, but the smaller the meson mass, the more flat the trend is and close to zero.
In addition, at the place closest to the threshold, the results extracted from the three vector meson photoproduction data are all minimal, indicating that the trace anomaly contribution to proton mass is only a tiny part.

Including the result from the differential cross section of $\phi$ photoproduction predicted by the two gluon exchange model, and the trace anomaly contribution obtained from the photoproduction experiment data of three light vector mesons are listed in Tab. \ref{tab:compare}. 
Here, we calculate the root mean square of all the results as ($4.36\pm0.40$)$\%$ to reflect the trace anomaly contribution to a general level, avoid the error caused by the uncertainty of the extraction process from different mesons, and make the result closer to the authentic value.

\section{Conclusion} \label{sec4}

In this paper, the trace anomaly contribution $M_{a}/M_{N}$ of the proton is extracted from the differential experimental cross sections data of the light vector mesons $\rho$, $\omega$ and $\phi$ photoproduction \cite{Dey:2014tfa,LEPS:2005hax,Strakovsky:2020uqs,rho-data,Barth:2003kv,Wu:2005wf} with the VMD model. Combined with the extraction $M_{a}/M_{N}$ of the cross section of $\phi$ photoproduction predicted at $R\in[0,0.66]$ by the two gluon exchange model, the root mean square of trace anomaly contribution is $(4.36\pm0.40$)$\%$, which is significantly smaller than that extracted from the heavy vector meson photoproduction. 

A new mass decomposition \cite{Metz:2020vxd} in 2020, also starting from the energy momentum tensor $T^{00}$, divides the proton mass into four terms. In their calculations, the trace anomaly term $M_{a}/M_{N}=0$ is considered, and the proton's mass is primarily composed of the other three parts.
All these results demonstrate that the trace anomaly contribution of the proton is tiny.
Although a lot of work has been done on the source of the proton mass, all of them focus on the difference between the normal and abnormal contributions, especially the trace anomaly contribution, how much it contributes to the proton mass is still the target of our further investigation.
Nevertheless, due to the lack of experimental data, the results at the closest threshold cannot be adequately described, and it is not sufficient to rely solely on the prediction of the two gluon exchange model. Therefore, acquiring more vector meson photoproduction data is necessary to investigate the source of proton mass further. In the future, the EIC facility \cite{Anderle:2021wcy,Accardi:2012qut} and JLab may have the opportunity to design more experiments to probe the proton's internal structure and explain the source of proton mass more reasonably.

\begin{acknowledgments}
This work is supported
by the National Natural Science Foundation of China under
Grant Nos. 12065014 and 12047501. We acknowledge the West Light Foundation of The Chinese Academy of Sciences, Grant No.
21JR7RA201.
\end{acknowledgments}


\begin{thebibliography}{99}

\bibitem{Polyakov:2018zvc}
M.~V.~Polyakov and P.~Schweitzer,
``Forces inside hadrons: pressure, surface tension, mechanical radius, and all that,''
Int. J. Mod. Phys. A \textbf{33}, 1830025 (2018).

\bibitem{Ji:1996ek}
X.~D.~Ji,
``Gauge-Invariant Decomposition of Nucleon Spin,''
Phys. Rev. Lett. \textbf{78}, 610-613 (1997).

\bibitem{Lorce:2017wkb}
C.~Lorc\'e, L.~Mantovani and B.~Pasquini,
``Spatial distribution of angular momentum inside the nucleon,''
Phys. Lett. B \textbf{776}, 38-47 (2018).

\bibitem{Lorce:2019zyw}
C.~Lorc\'e,
``Energy, angular momentum and pressure force distributions inside nucleons,''
J. Phys. Conf. Ser. \textbf{1643}, 012190 (2020).

\bibitem{Granados:2019zjw}
C.~Granados and C.~Weiss,
``Partonic angular momentum in the nucleon's chiral periphery,''
Phys. Lett. B \textbf{797}, 134847 (2019).

\bibitem{Wang:2022uch}
X.~Y.~Wang, C.~Dong and Q.~Wang,
``Mass radius and mechanical properties of the proton via strange $\phi$ meson photoproduction,''
[arXiv:2206.11644 [nucl-th]].

\bibitem{Kharzeev:2021qkd}
D.~E.~Kharzeev,
``Mass radius of the proton,''
Phys. Rev. D \textbf{104}, 054015 (2021).

\bibitem{Kou:2021qdc}
W.~Kou, R.~Wang and X.~Chen,
``Determination of the Gluonic D-term and Mechanical Radii of Proton from Experimental data,''
[arXiv:2104.12962 [hep-ph]].

\bibitem{Polyakov:2002wz}
M.~V.~Polyakov and A.~G.~Shuvaev,
``On'dual' parametrizations of generalized parton distributions,''
[arXiv:hep-ph/0207153 [hep-ph]].

\bibitem{Lorce:2018egm}
C.~Lorc\'e, H.~Moutarde and A.~P.~Trawi\'nski,
``Revisiting the mechanical properties of the nucleon,''
Eur. Phys. J. C \textbf{79}, 89 (2019).

\bibitem{Wang:2022vhr}
X.~Y.~Wang, F.~Zeng and Q.~Wang,
``Systematic analysis of the proton mass radius based on photoproduction of vector charmoniums,''
Phys. Rev. D \textbf{105}, 096033 (2022).

\bibitem{Metz:2020vxd}
A.~Metz, B.~Pasquini and S.~Rodini,
``Revisiting the proton mass decomposition,''
Phys. Rev. D \textbf{102}, 114042 (2020).

\bibitem{Ji:1995sv}
X.~D.~Ji,
``Breakup of hadron masses and energy - momentum tensor of QCD,''
Phys. Rev. D \textbf{52}, 271-281 (1995).

\bibitem{Ji:1994av}
X.~D.~Ji,
``A QCD analysis of the mass structure of the nucleon,''
Phys. Rev. Lett. \textbf{74}, 1071-1074 (1995).

\bibitem{Lorce:2017xzd}
C.~Lorc\'e,
``On the hadron mass decomposition,''
Eur. Phys. J. C \textbf{78}, 120 (2018).

\bibitem{Hatta:2018sqd}
Y.~Hatta, A.~Rajan and K.~Tanaka,
``Quark and gluon contributions to the QCD trace anomaly,''
JHEP \textbf{12}, 008 (2018).

\bibitem{DUNE:2021tad}
A.~Abed Abud \textit{et al.} [DUNE],
``Deep Underground Neutrino Experiment (DUNE) Near Detector Conceptual Design Report,''
Instruments \textbf{5},, 31 (2021).

\bibitem{Wang:2019mza}
R.~Wang, J.~Evslin and X.~Chen,
``The origin of proton mass from J/${\Psi }$ photo-production data,''
Eur. Phys. J. C \textbf{80}, 507 (2020).

\bibitem{Kou:2021bez}
W.~Kou, R.~Wang and X.~Chen,
``Extraction of proton trace anomaly energy from near-threshold $\phi $ and $J/\psi $ photo-productions,''
Eur. Phys. J. A \textbf{58}, 155 (2022).

\bibitem{LEPS:2005hax}
T.~Mibe \textit{et al.} [LEPS],
``Diffractive phi-meson photoproduction on proton near threshold,''
Phys. Rev. Lett. \textbf{95}, 182001 (2005).

\bibitem{Barth:2003bq}
J.~Barth, W.~Braun, J.~Ernst, K.~H.~Glander, J.~Hannappel, N.~Jopen, F.~J.~Klein, F.~Klein, H.~Kalinowsky and E.~Klempt, \textit{et al.}
``Low-energy photoproduction of Phi mesons,''
Eur. Phys. J. A \textbf{17}, 269-274 (2003).

\bibitem{GlueX:2019mkq}
A.~Ali \textit{et al.} [GlueX],
``First Measurement of Near-Threshold J/\ensuremath{\psi} Exclusive Photoproduction off the Proton,''
Phys. Rev. Lett. \textbf{123}, 072001 (2019).

\bibitem{Duran:2022xag}
B.~Duran, Z.~E.~Meziani, S.~Joosten, M.~K.~Jones, S.~Prasad, C.~Peng, W.~Armstrong, H.~Atac, E.~Chudakov and H.~Bhatt, \textit{et al.}
``When Color meets Gravity; Near-Threshold Exclusive $J/\psi$ Photoproduction on the Proton,''
[arXiv:2207.05212 [nucl-ex]].

\bibitem{Dey:2014tfa}
B.~Dey \textit{et al.} [CLAS],
``Data analysis techniques, differential cross sections, and spin density matrix elements for the reaction $\gamma p \rightarrow \phi p$,''
Phys. Rev. C \textbf{89}, 055208 (2014).

\bibitem{Strakovsky:2020uqs}
I.~I.~Strakovsky, L.~Pentchev and A.~Titov,
``Comparative analysis of $\omega p$, $\phi p$, and $J/\psi p$ scattering lengths from A2, CLAS, and GlueX threshold measurements,''
Phys. Rev. C \textbf{101}, 045201 (2020).

\bibitem{Barth:2003kv}
J.~Barth, W.~Braun, J.~Ernst, K.~H.~Glander, J.~Hannappel, N.~Jopen, H.~Kalinowsky, F.~J.~Klein, F.~Klein and E.~Klempt, \textit{et al.}
``Low-energy of photoproduciton of omega-mesons,''
Eur. Phys. J. A \textbf{18}, 117-127 (2003).

\bibitem{rho-data}
SAPHIR Collaboration, F. J. Klein et al.,
``$\pi$N Newsletter 14, 141 (1998),''
J. Klein, Ph. D. Thesis, Bonn Univ. (1996).

\bibitem{Wu:2005wf}
C.~Wu, J.~Barth, W.~Braun, J.~Ernst, K.~H.~Glander, J.~Hannappel, N.~Jopen, H.~Kalinowsky, F.~J.~Klein and F.~Klein, \textit{et al.}
``Photoproduction of rho0 mesons and Delta-baryons in the reaction gamma p ---\ensuremath{>} p pi+ pi- at energies up to s**(1/2) = 2.6-GeV,''
Eur. Phys. J. A \textbf{23}, 317-344 (2005)

\bibitem{Buras:1979yt}
A.~J.~Buras,
``Asymptotic Freedom in Deep Inelastic Processes in the Leading Order and Beyond,''
Rev. Mod. Phys. \textbf{52}, 199 (1980).

\bibitem{Baikov:2014qja}
P.~A.~Baikov, K.~G.~Chetyrkin and J.~H.~K\"uhn,
``Quark Mass and Field Anomalous Dimensions to ${\cal O}(\alpha_s^5)$,''
JHEP \textbf{10}, 076 (2014).

\bibitem{Kharzeev:1998bz}
D.~Kharzeev, H.~Satz, A.~Syamtomov and G.~Zinovjev,
``J / psi photoproduction and the gluon structure of the nucleon,''
Eur. Phys. J. C \textbf{9}, 459-462 (1999).

\bibitem{Kharzeev:1995ij}
D.~Kharzeev,
``Quarkonium interactions in QCD,''
Proc. Int. Sch. Phys. Fermi \textbf{130}, 105-131 (1996).

\bibitem{Kharzeev:1996tw}
D.~Kharzeev, H.~Satz, A.~Syamtomov and G.~Zinovev,
``On the sum rule approach to quarkonium - hadron interactions,''
Phys. Lett. B \textbf{389}, 595-599 (1996).

\bibitem{Pentchev:2020kao}
L.~Pentchev and I.~I.~Strakovsky,
``$J/\psi$-$p$ Scattering Length from the Total and Differential Photoproduction Cross Sections,''
Eur. Phys. J. A \textbf{57}, 56 (2021).

\bibitem{Goloskokov:2007zb}
S.~V.~Goloskokov,
``Electroproduction of Light Vector Mesons,''
[arXiv:0712.3968 [hep-ph]].

\bibitem{Gluck:1998xa}
M.~Gl\"uck, E.~Reya and A.~Vogt,
``Dynamical parton distributions revisited,''
Eur. Phys. J. C \textbf{5}, 461-470 (1998).

\bibitem{Ball:2011uy}
R.~D.~Ball \textit{et al.} [NNPDF],
``Unbiased global determination of parton distributions and their uncertainties at NNLO and at LO,''
Nucl. Phys. B \textbf{855}, 153-221 (2012).

\bibitem{Owens:2012bv}
J.~F.~Owens, A.~Accardi and W.~Melnitchouk,
``Global parton distributions with nuclear and finite-$Q^2$ corrections,''
Phys. Rev. D \textbf{87}, 094012 (2013).

\bibitem{Accardi:2016qay}
A.~Accardi, L.~T.~Brady, W.~Melnitchouk, J.~F.~Owens and N.~Sato,
``Constraints on large-$x$ parton distributions from new weak boson production and deep-inelastic scattering data,''
Phys. Rev. D \textbf{93}, 114017 (2016).

\bibitem{Wang:2016sfq}
R.~Wang and X.~Chen,
``Dynamical parton distributions from DGLAP equations with nonlinear corrections,''
Chin. Phys. C \textbf{41}, 053103 (2017).


\bibitem{Zeng:2020coc}
F.~Zeng, X.~Y.~Wang, L.~Zhang, Y.~P.~Xie, R.~Wang and X.~Chen,
``Near-threshold photoproduction of $J/\psi$ in two-gluon exchange model,''
Eur. Phys. J. C \textbf{80}, 1027 (2020).

\bibitem{Pumplin:2002vw}
J.~Pumplin, D.~R.~Stump, J.~Huston, H.~L.~Lai, P.~M.~Nadolsky and W.~K.~Tung,
``New generation of parton distributions with uncertainties from global QCD analysis,''
JHEP \textbf{07}, 012 (2002).

\bibitem{Sibirtsev:2004ca}
A.~Sibirtsev, S.~Krewald and A.~W.~Thomas,
``Systematic analysis of charmonium photoproduction,''
J. Phys. G \textbf{30}, 1427-1444 (2004).

\bibitem{peskin}
M.E. Peskin and D.V. Schroeder,
``An Introduction to Quantum Field Theory (Addison-Wesley,1995) 842.''

\bibitem{Wang:2022tuc}
X.~Y.~Wang, C.~Dong and Q.~Wang,
``Analysis of the interaction between of $\phi$ meson and nucleus,''
[arXiv:2208.10289 [nucl-th]].

\bibitem{Mathieu:2018xyc}
V.~Mathieu \textit{et al.} [JPAC],
``Vector Meson Photoproduction with a Linearly Polarized Beam,''
Phys. Rev. D \textbf{97}, 094003 (2018).

\bibitem{Zhao:1998rt}
Q.~Zhao, Z.~p.~Li and C.~Bennhold,
``Omega and rho photoproduction with an effective quark model Lagrangian,''
Phys. Lett. B \textbf{436}, 42-48 (1998).

\bibitem{Ganbold:2010bu}
G.~Ganbold,
``QCD Effective Coupling in the Infrared Region,''
Phys. Rev. D \textbf{81}, 094008 (2010).

\bibitem{Accardi:2012qut}
A.~Accardi, J.~L.~Albacete, M.~Anselmino, N.~Armesto, E.~C.~Aschenauer, A.~Bacchetta, D.~Boer, W.~K.~Brooks, T.~Burton and N.~B.~Chang, \textit{et al.}
``Electron Ion Collider: The Next QCD Frontier: Understanding the glue that binds us all,''
Eur. Phys. J. A \textbf{52}, 268 (2016).

\bibitem{Anderle:2021wcy}
D.~P.~Anderle, V.~Bertone, X.~Cao, L.~Chang, N.~Chang, G.~Chen, X.~Chen, Z.~Chen, Z.~Cui and L.~Dai, \textit{et al.}
``Electron-ion collider in China,''
Front. Phys. (Beijing) \textbf{16}, 64701 (2021).


\end{thebibliography}
\end{document}